# A Simple, Flexible Technique for RF Cavity Wake-Field Calculations

Brian J. Vaughn*
*Fermi National Accelerator Laboratory, Batavia, Illinois 60510*

It is typical in the accelerator field to model machine components, especially RF cavities, as parallel RLC resonators. To properly model wake-fields, knowledge of the time-domain voltage resulting from beam excitation is often necessary. While analytical and quasi-analytical expressions are available to accomplish this for common bunch distributions such as the Gaussian, analogous results for less standard distributions can be difficult or computationally-taxing to obtain using direct methods, which opens the door for the development of a more generalized technique. In this paper, a formulation is created that allows for the simple computation of the time-domain voltage waveform of an RLC resonator. The formulation uses the Cauchy Residue Theorem to extract the convolution result from the Fourier Domain, and if current distribution Fourier Transform has no poles, knowledge of its value is only required at one specific evaluation point. This greatly simplifies the computation of the time domain voltage for a large amount of bunch distributions both common and uncommon. Accuracy considerations for this technique and the approximation of accelerator components as RLC resonators are also discussed, resulting the development of a figure of merit for quantifying the robustness of this type of approximation.

1. **Introduction**

It is well-established in accelerator science that machine structures (beam pipes, cavities, etc.) are subject to electromagnetic excitation from the beam bunch currents passing through them. This fact has given rise to the concept of wake-fields as well as the study and simulation of the perturbative effects and beam instabilities these fields present [1]-[3]. The importance of forming predictive and diagnostic models of the instabilities within machines has resulted in a high amount of attention given to the mathematical treatment of beam-machine interactions. For example, the Robinson instability formulation uses machine longitudinal impedance curves to quantify longitudinal emittance growth or damping due to cavity-beam excitations [1]. While this formulation is popular and useful, it is primarily a frequency-domain technique and assumes a machine with uniform-fill (some necessary techniques used in this formulation, such as the Poisson Summation Theorem, are far less conducive to non-uniform fill scenarios, which typically require numerical treatment [4]). On the other hand, it is often useful to understand individualized longitudinal bunch perturbations on a turn-by-turn basis, which requires time-domain knowledge. For example, it is standard to model the impedance of an RF accelerating cavity, or even the impedance of an entire beamline, in terms of parallel RLC resonators [5]-[8]. The time-domain voltage over such resonators may be computed by convolving an equivalent bunch current distribution with the RLC resonator impulse response [5], [9]. That is,

$$V(t) = I_b(t) * z(t) = \int_{-\infty}^{\infty} I_b(t - \tau) z(\tau) d\tau, \tag{1a}$$

$$z(t) = \frac{\omega_r R_s e^{\frac{-\omega_r t}{2Q}}}{Q \sqrt{4 - \frac{1}{Q^2}}} \left[ \sqrt{4 - \frac{1}{Q^2}} \cos \frac{\omega_r t}{2} \sqrt{4 - \frac{1}{Q^2}} - \frac{1}{Q} \sin \frac{\omega_r t}{2} \sqrt{4 - \frac{1}{Q^2}} \right], \tag{1b}$$

where $V(t)$ is the resonator voltage, $I_b(t)$ is the equivalent bunch current, $z(t)$ is the resonator impulse response, $\omega_r$ is the angular resonant frequency, $R_s$ is the shunt resistance, and $Q$ is the quality factor. Fig. 1 shows the equivalent resonator circuit under bunch current excitation. Fig. 1 demonstrates the equivalent circuit for only one resonant mode, as well as the circuit for multiple modes exhibited by the same structure, which is the more general case. The multi-modal case may be modeled as a circuit with $H$ RLC resonators connected in series, where $H$ is the number of modes in question. It is a series connection instead of parallel since several RLC resonators connected in parallel would not display a multi-modal response. Instead, the parallel combination collapses to a single-mode circuit with a modified resonant frequency and/or Q. Since the equivalent resonators are series-connected, they all carry the same current $I_b(t)$. As such, by Kirchoff's Voltage Law, the total voltage developed over the full multi-modal circuit may be taken as the superposition of the voltages over each constituent resonator, which may be computed via eqn. (1a). Given this, we will begin our analysis with the single-mode case in and then expand to the multi-mode case by superposition.

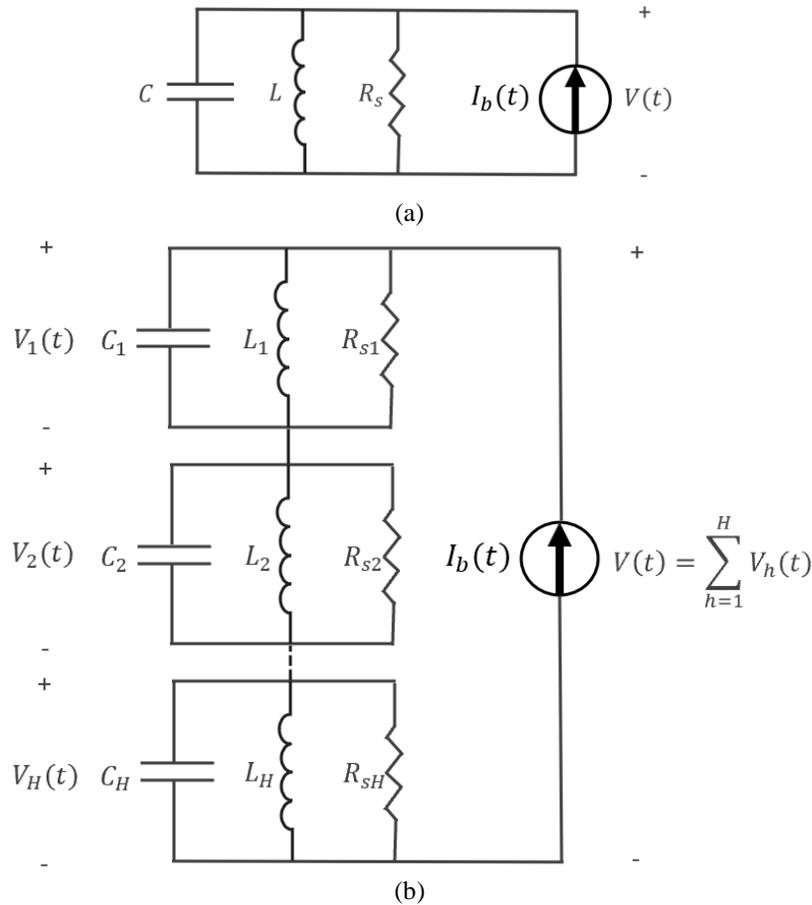

**FIG. 1.** Parallel RLC resonator circuit, which may be used to model beamline components or a single mode of an accelerating cavity. $L$ and $C$ are the equivalent inductance and capacitance respectively. The figure shows the circuit for a single mode (a) as well as multiple modes in superposition (b).

While the form of $V(t)$ has been computed for certain common distributions such as the Gaussian, evaluating the convolution integral for eqn. (1a) can be quite cumbersome in general when done analytically or time-consuming if the convolution integral is computed purely numerically in a direct manner. In fact, commonly accepted expressions for the wake function of a resonator due to a Gaussian bunch (which is functionally equivalent to the voltage in eqn. (1a)) require use of the Error Function and are thus not closed-form, adding unnecessary computations to already taxing processes such as large turn-by-turn simulations (see the table at the end of Section 3.2.3 of [10]). In this paper, we will simplify this evaluation by developing a versatile analytical technique for solving the convolution of a bunch current distribution and the parallel RLC resonator impulse response. This will be done by invoking the convolution property of the Fourier Transform (FT) and recasting the inverse FT integral such that it may be easily computed using the Cauchy Residue Theorem. Furthermore, if the bunch current distribution is holomorphic, the value of the bunch current FT needs only to be known at one specific frequency point per resonance in order to carry out the technique. This results in a powerful, compact formulation that may be used to represent the time-domain voltage over an RLC resonator due to a wide variety of bunch excitations. To the author's knowledge, no such compact method for this mathematical task has been reported. This paper's structure is as follows: Sections II and III will develop the formulation beginning in the Fourier Domain, Section IV will discuss how effective bunch excitation distributions are perturbed by finite transit time through accelerator components, Section V will discuss the accuracy considerations of applying the given formulation, and Section VI contains examples of the formulation applied to non-Gaussian current distributions relevant to accelerator science.

## 2. Formulation Development

We begin with the well-documented impedance of a single-mode parallel RLC resonator shown in Fig. 1:

$$Z(\omega) = \frac{R_s}{1 - jQ\left(\frac{\omega_r}{\omega} - \frac{\omega}{\omega_r}\right)}. \tag{2}$$

Note that the engineer's definition of the FT is used herein (hence the $-j$ in the denominator) and that $j$ is the imaginary number. By rearranging terms, we may express eqn. (2) in the following form:

$$Z(\omega) = \frac{R_s\left(\frac{\omega}{\omega_r}\right)}{\frac{\omega}{\omega_r} - jQ\left(1 - \frac{\omega^2}{\omega_r^2}\right)}. \tag{3}$$

Next, we may make the following substitution:

$$Z(u) = \frac{R_s u}{u - jQ(1 - u^2)}, \tag{4a}$$

$$u \equiv \frac{\omega}{\omega_r}. \tag{4b}$$

The denominator of eqn. (4a) is a quadratic equation in $u$, so we may rearrange terms and express the denominator in terms of the roots of this quadratic equation:

$$Z(u) = \frac{1}{jQ} \frac{R_s u}{(u - r^+)(u - r^-)}, \tag{5a}$$

$$r^\pm = \frac{1}{2}\left(j/Q \pm \sqrt{4 - \frac{1}{Q^2}}\right). \tag{5b}$$

Eqn. (5a) demonstrates that the RLC impedance expression is a meromorphic function with two complex poles at $r^+$ and $r^-$. We will now invoke the FT of the bunch profile as seen by the resonator and perform the same change of variable to express it as a function of $u$. We will denote this FT as $F_b(u)$. For now, let us only consider a holomorphic $F_b(u)$ such that the bunch FT does not contain any singularities (we will relax this condition later). From the FT convolution-product property, we may express the FT of the bunch-induced resonator voltage as:

$$F_v(u) = \frac{1}{jQ} \frac{R_s u F_b(u)}{(u - r^+)(u - r^-)}. \tag{6}$$

Our objective is to express the bunch-induced voltage excitation in the time domain by taking the inverse FT of eqn. (6):

$$\check{F}^{-1}(F_v(u)) = \frac{\omega_r R_s}{j 2\pi Q} \int_{-\infty}^{\infty} \frac{u F_b(u)}{(u - r^+)(u - r^-)} e^{j\omega_r u t} du. \tag{7}$$

Note a change of variable from $\omega$ to $u$ has been conducted in the inverse FT integral. Since the poles in the above expression are complex, we may now invoke the Cauchy Residue Theorem to evaluate the real-line improper inverse FT integral. The theorem states that the closed contour integral of a loop containing the poles of a meromorphic function in the complex plane is equal to the sum of the residues corresponding to those poles multiplied by $2\pi j$. That is:

$$\oint_C f(u) du = 2\pi j \sum_n Res(r_n). \tag{8}$$

In the complex $u$ plane, then, we may define a semicircular contour with one branch spanning the entire real line as per Fig. 2. To compute the residues for each pole, we may define four functions $g^+(u) = \frac{u F_b(u) e^{j\omega_r u t}}{u - r^-}$, $g^-(u) = \frac{u F_b(u) e^{j\omega_r u t}}{u - r^+}$, $h^+(u) = u - r^+$, and $h^-(u) = u - r^-$. The residues may then be easily calculated via the following formula [12]:

$$Res(r^\pm) = \frac{g^\pm(r^\pm)}{h^{\pm\prime}(r^\pm)} = \frac{\pm r^\pm F_b(r^\pm) e^{j\omega_r r^\pm t}}{r^+ - r^-}. \tag{9}$$

Therefore,

$$\int_{-\infty}^{\infty} \frac{u F_b(u)}{(u - r^+)(u - r^-)} e^{j\omega_r u t} du$$

$$= 2\pi j \left(\frac{r^+ F_b(r^+) e^{j\omega_r r^+ t} - r^- F_b(r^-) e^{j\omega_r r^- t}}{r^+ - r^-}\right) - \int_{C_{outer}} \frac{u F_b(u)}{(u - r^+)(u - r^-)} e^{j\omega_r u t} du. \tag{10}$$

The $C_{outer}$ contour integral represents the circular portion of the Fig. 2 semicircular contour that traverses the path between $+\infty$ and $-\infty$ through the complex space where $Im(u)$ is not necessarily 0 and $|u| \to \infty$. This integral vanishes under the proper conditions. A rigorous justification for this claim may be obtained by invoking the lemma found and proved in Sec. 43 of [11], which states that for a complex-valued function $w(s)$ defined on some interval $a \leq s \leq b$,

$$\left|\int_a^b w(s)ds\right| \leq \int_a^b |w(s)|ds. \tag{11a}$$

We may connect this lemma to our use case by executing a change of variables in the RHS integral of eqn. (10):

$$u = R_0 e^{j\theta} \Rightarrow \int_{C_{outer}} \frac{uF_b(u)}{(u-r^+)(u-r^-)} e^{j\omega_r ut} du = \int_0^\pi \lim_{R_0 \to \infty} \frac{R_0 e^{j\theta} F_b(R_0 e^{j\theta})}{(R_0 e^{j\theta} - r^+)(R_0 e^{j\theta} - r^-)} jR_0 e^{j\theta} e^{j\omega_r R_0 e^{j\theta} t} d\theta$$

$$= \int_0^\pi \lim_{R_0 \to \infty} jF_b(R_0 e^{j\theta}) e^{j\omega_r R_0 e^{j\theta} t} d\theta, \tag{11b}$$

where $R_0, \theta \in \mathbb{R}$. Expanding the exponential term yields

$$e^{j\omega_r R_0 e^{j\theta} t} = e^{jRe(e^{j\theta})\omega_r R_0 t} e^{-Im(e^{j\theta})\omega_r R_0 t}. \tag{11c}$$

Since $0 \leq \theta \leq \pi$, $Im(e^{j\theta}) \geq 0$, so the exponential term vanishes as $R_0 \to \infty$ for $t > 0$. Since the conditions of this technique stipulate that $F_b$ contains no singularities on the contour with $R_0 = \infty$ (we will retain this requirement even when we relax the holomorphism constraint), the integrand of equation (11c) vanishes over the entire contour. Therefore, by eqn. (11a),

$$\left|\int_{C_{outer}} \frac{uF_b(u)}{(u-r^+)(u-r^-)} e^{j\omega_r ut} du\right| = \left|\int_0^\pi \lim_{R_0 \to \infty} jF_b(R_0 e^{j\theta}) e^{j\omega_r R_0 e^{j\theta} t} d\theta\right|$$

$$\leq \int_0^\pi \left|\lim_{R_0 \to \infty} jF_b(R_0 e^{j\theta}) e^{j\omega_r R_0 e^{j\theta} t}\right| d\theta = 0. \tag{12}$$

Note again that this proof only applies for evaluations of the Inverse Fourier Transform for $t > 0$, otherwise the contour integral can't be guaranteed to vanish unless $F_b(R_0 e^{j\theta})$ shrinks more quickly than the exponential term in the integrand grows. For accelerator applications, this restriction does not result in a loss of utility since we are only interested in the excited voltage insofar as it affects bunches that arrive after the excitation bunch, which only necessitates evaluation at points where $t > 0$ assuming that the $t = 0$ reference point is properly chosen relative to the excitation bunch current arrival. Moreover, bunch self-excitation is described by the Fundamental Theorem of Beam Loading, so the technique discussed here is not needed for that case.

We therefore obtain the following result for the cavity time-domain voltage response, which we may immediately expand to the multi-mode case, denoting this total voltage as $V^{MM}(t)$.

$$V(t) = \breve{F}^{-1}(F_v(u)) = \frac{\omega_r R_s}{Q}\left(\frac{1}{r^+ - r^-}\right)\left(r^+ F_b(r^+) e^{j\omega_r r^+ t} - r^- F_b(r^-) e^{j\omega_r r^- t}\right), \tag{13a}$$

$$V^{MM}(t) = \sum_{h=1}^H \breve{F}^{-1}(F_{vh}(u)) = \sum_{h=1}^H \frac{\omega_{rh} R_{sh}}{Q_h}\left(\frac{1}{r_h^+ - r_h^-}\right)\left(r_h^+ F_b(r_h^+) e^{j\omega_{rh} r_h^+ t} - r_h^- F_b(r_h^-) e^{j\omega_{rh} r_h^- t}\right), \tag{13b}$$

where $H$ is the total number of modes under consideration and any value with an $h$ subscript is that value computed from the parameters of the $h^{th}$ mode. We will discuss the appropriate number of modes to consider in Section V. Note that since $r_h^+$ and $r_h^-$ have the same imaginary part and real parts that are only different in sign, $F_b(r_h^+)$ and $F_b(r_h^-)$ will always be complex conjugates of each other. In conclusion, eqn. (13) may be used to find the time-domain voltage response of a resonator mode induced by a bunch current profile with a holomorphic Fourier transform. This method is computationally simple and fast when compared to direct numerical convolution or complicated analytical evaluation. Note that, while this expression contains complex terms, if the equation inputs are physical, the end result will be purely real. It is also worth emphasizing that eqn. (13) only requires the value of the bunch profile FT to be evaluated at one point; $r^+$ or $r^-$. This is a powerful property, as it allows for the equation to be used even if the FT of the bunch profile can only be obtained numerically. Referring again to the Gaussian bunch, which has a well-defined analytical Fourier Transform, we see that the voltage may be expressed in a purely analytical manner using eqn. (13), eliminating the need for Error Function computation. Finally, it should be restated that, in order to account for resonator multi-modality, eqn. (13) may be evaluated for each individual mode and the total voltage distribution can be found by summing the individual evaluations together.

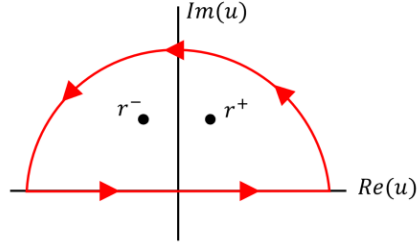

**FIG. 2.** Semicircular contour in complex $u$ plane that encloses the two integrand poles. This contour may be used to evaluate an improper integral on the real line if the semicircle radius approaches infinity.

3. **Non-Holomorphic Bunch Transform**

We will now consider cases where the bunch FT is meromorphic with $N$ isolated poles and has the form

$$F_b(u) = \frac{G_b(u)}{\prod_{n=1}^{N}(u - r_n)^{k_n}}, \{r_n \in \mathbb{C}\}, \{k_n \in \mathbb{Z} \mid k_n \geq 1\}. \tag{14}$$

Where $G_b(u)$ is a nonzero holomorphic function. As before, let $F_b$ be finite for all arguments where $|u| = \infty$. The integral we now wish to evaluate is

$$\int_{-\infty}^{\infty} \frac{uG_b(u)}{(u - r^+)(u - r^-)\prod_{n=1}^{N}(u - r_n)^{k_n}} e^{j\omega_r ut} du. \tag{15}$$

In this case, the residue summation of eqn. (8) must now include contributions from $F_b(u)$ for all singularities that lie in the defined semi-circular contour. If any singularity lies on the real axis, its residue contribution is multiplied by ½ since we may define an infinitesimally small half-circular arc around this singularity that converges to the real line in the limit and constitutes half of the contour integral for that point [12]. Since we are allowing the bunch distribution pole orders to exceed 1, we will adopt a more general form for residue computation [12]:

$$Res_m(r_m) = \frac{\varphi_m^{(k_m-1)}(r_m)}{(k_m - 1)!}, \tag{16a}$$

$$\varphi_m(u) = (u - r_m)^{k_m} \frac{uG_b(u)}{(u - r^+)(u - r^-)\prod_{n=1}^{N}(u - r_n)^{k_n}} e^{j\omega_r ut}, \tag{16b}$$

where $r_m$ is the position of the $m^{th}$ pole and $k_m$ is the pole's order. Using the same argument as that used for eqn. (12), it is clear that the integrand shown in eqn. (15) also vanishes as $|u| \to \infty$. As such, we may evaluate the inverse Fourier transform and expand to the multi-mode case as

$$V(t) = \breve{F}^{-1}(F_v(u)) = \frac{\omega_r R_s}{Q}\left(\sum_{m=1}^{M}\frac{\varphi_m^{(k_m-1)}(r_m)}{(k_m - 1)!} + \frac{1}{2}\sum_{p=M+1}^{M+P}\frac{\varphi_p^{(k_p-1)}(r_p)}{(k_p - 1)!}\right), \tag{17a}$$

$$V^{MM}(t) = \sum_{h=1}^{H}\breve{F}^{-1}(F_{vh}(u)) = \sum_{h=1}^{H}\frac{\omega_{rh}R_{sh}}{Q_h}\left(\sum_{m=1}^{M}\frac{\varphi_{mh}^{(k_{mh}-1)}(r_{mh})}{(k_{mh} - 1)!} + \frac{1}{2}\sum_{p=M+1}^{M+P}\frac{\varphi_{ph}^{(k_{ph}-1)}(r_{ph})}{(k_{ph} - 1)!}\right), \tag{17b}$$

where $M$ is the number of unique singularities that do not lie on the real axis (including those introduced by the resonator impedance, i.e., $r^+$ and $r^-$) and $P$ is the number of singularities that do, with the indices chosen such that singularities 1 through $M$ are of the former category and singularities $M + 1$ through $P$ are of the latter. Note that $M + P$ is not necessarily equal to the total number of unique poles for both the resonator and bunch transform if the bunch transform includes singularities in the lower half of the complex plane. In this case, the poles/orders of these lower half-plane singularities would be indexed at numbers greater than $M + P$. Alternatively, if there are fewer lower half-plane poles than upper half-plane poles, it may be more convenient to use a lower half-plane contour to evaluate the integral. If this is desired, eqn. (17) may be modified by changing the bounds of the first pole summation from 1 to $M$ to $M + P + 1$ to $H$ (where $H$ is the total number of unique poles for the integrand) and multiplying the entire expression by -1 since the lower half-plane contour must be clockwise to align with the improper real-line integral.

## 4. Bunch Excitation Stretching and Scaling

For the most accurate results, care should be taken when projecting a given bunch profile onto the excitation current distribution of an accelerator component, namely, a cavity. This is because, the cavity "sees" a version of the bunch current profile that is perturbed by the finite length of the accelerating gap. The result is an effective current excitation that is stretched out by the time the bunch spends traversing the gap and scaled accordingly. If we define some characteristic width $t_w$ that describes the width of the bunch current distribution (e.g., the two-sided standard deviation $2\sigma$ for a Gaussian profile), then we may deduce that a bunch current passing through a cavity gap of width $g$ at velocity $v$ will have its characteristic width extended by $\frac{g}{v}$. This is because each charge within the bunch takes a time of $\frac{g}{v}$ to traverse the gap. This is visualized in Fig. 3. As such, if we define the nominal bunch current distribution as $i_b(t)$, then the gap-perturbed current will be of the form $i_b\left(t\frac{t_w}{t_w+\frac{g}{v}}\right)$. Note that for this discussion, $t=0$ is defined relative to the center of the gap. Additionally, the current distribution with respect to the cavity gap must be scaled such that the total bunch charge is conserved. The total bunch charge may be expressed as

$$Q_b = \int_{-\infty}^{\infty} i_b(t)\, dt. \tag{18}$$

We may then define a scaling factor $Q_{scale}$ such that

$$Q_{scale} = \frac{Q_b}{\int_{-\infty}^{\infty} i_b\left(t\dfrac{t_w}{t_w+\dfrac{g}{v}}\right) dt}. \tag{19}$$

Therefore, the bunch current profile as seen by the cavity can be expressed as

$$i_b^p(t) = Q_{scale}\, i_b\left(t\frac{t_w}{t_w+\dfrac{g}{v}}\right). \tag{20}$$

It is this current profile that should be used to find $F_h(u)$ from the previous section.

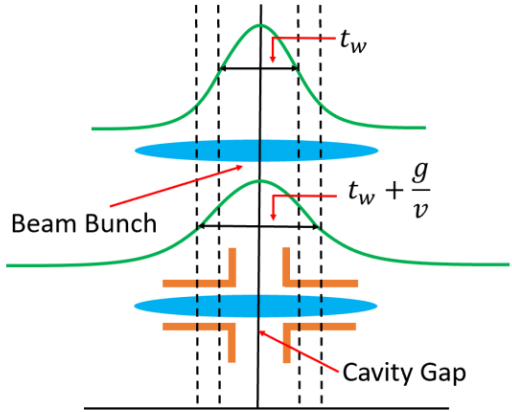

**FIG. 3.** Bunch current profile perturbation. The top curve shows the nominal bunch current distribution, and the bottom curve shows the distribution stretched out by the finite cavity gap width.

## 5. Accuracy

In order to obtain accurate results using the above formulation, it is important to define its limitations and conditions of appropriateness. In this section, we will consider several aspects of the technique that are linked to these considerations.

### A. Number of Modes

In the case of an RF cavity, the number of modal solutions is functionally infinite. Clearly, then, one must decide which modes to consider and which to neglect, and care must be taken when doing so if errors are to be avoided. When designing an RF cavity, it is typical to map out the modal parameters well beyond those of the fundamental mode, namely, the resonant frequency, Q, and shunt resistance. It is also typical to apply Higher-Order-Mode (HOM) dampers primarily to reduce the Q of unwanted modes such that they will decay before they have the opportunity to perturb the beam. As $Q \to 0$, for example, $r^+ \to j\infty$ and $r^- \to 0$, so the negative residue term in eqn. (13) becomes null and the positive residue term vanishes infinitely quickly for any $t > 0$. As such, modes associated with a low Q may be neglected. Other factors that impact HOM effects are the R/Q values (higher R/Q leads to larger voltage excitation) and the frequency content of the bunch current FT around the resonant frequencies of the HOM's. For example, as $Q \to \infty$, $r^\pm \to \pm 1$. If $\frac{\omega_r R_s}{Q} F_v(u) \approx 0$ for $u = \pm 1$, this mode may also be safely neglected (note $R_s$ would also approach $\infty$ in this case at the same rate as Q, meaning R/Q would still be finite). Given these factors, we may recontextualize the multi-mode voltage excitation without loss of generality as

$$V^{MM}(t) = \sum_{h=1}^{H} V_h(t) + \sum_{l=H+1}^{\infty} V_l(t), \tag{21}$$

where $H$ is the minimum number of modes such that

$$\sum_{l=H+1}^{\infty} V_l(t) \ll \sum_{h=1}^{H} V_h(t) => V^{MM}(t) \approx \sum_{h=1}^{H} V_h(t), t \geq t_0, \tag{22}$$

where $t_0$ is some minimum time before which the value of the voltage excitation is not relevant to the desired study. Typically, $t_0$ will be the arrival time of the bunch immediately following the excitation bunch.

**B. Bunch Length**

Here, we note that, if intra-bunch effects such as space charge and long-bunch self-excitation (which are beyond the scope of this paper) are neglected, then the accuracy of the technique detailed in this paper is largely independent of bunch length. Furthermore, the result of these effects, if accounted for, would be the perturbation of the bunch current distribution as it passes through the component being modelled as a resonator. The impulse response, on the other hand, is fundamentally a quantity that exists independently of the excitation, which, if accurate, only comes into play when evaluating the convolution integral and deciding how many modes to consider during the computation. This point is reinforced by Fig. 1 of [5], which details the differences in numerical convolution calculations for two Gaussian bunches, one of which is short (2 cm) and the other long (2 m). For comparison, both cases use the analytical evaluation for the Gaussian wake-function given in [10]. Since the validity of the computations in that work are taken to be sufficient for majorly disparate bunch lengths, we may conclude that the same is true for the techniques discussed here.

**C. Resonator Approximation of Accelerator Components**

As alluded to above, it is common to compute the impedance of one or more non-cavity accelerator components and map it to a broadband RLC resonator with a frequency response that approximates that of the computed impedance within the frequency range of interest. For example, if the impedance of a component is inductive within a certain frequency range, an appropriate choice could be to model the component as a parallel RLC resonator with a resonant frequency much higher than the range of interest and a Q and $R_s$ that allow for proper frequency dependence fit. Of course, the quality of this fit will affect the accuracy of the model and divergence outside the desired range is inevitable unless the component behaves like an RLC resonator natively. This divergence is an important consideration since the formulation above depends on the evaluation of $F_v(u)$ at discrete points which must be accurately represented by the model. That said, if $F_b(u) \approx 0$ in the regions where the modelled and true responses diverge, the accuracy of the inverse Fourier integral will be minimally affected. With these effects in mind, we wish to quantify the robustness of a given resonator approximation. Consider an accelerator component with an impedance given by $Z_c(\omega)$ that is explicitly known on some finite frequency range $[\omega_a, \omega_b]$. Assuming we aim to map this impedance to some equivalent resonator, we define the following squared-error parameter:

$$\varepsilon^2 = \int_{-\infty}^{\infty} |[Z_r(\omega) - Z_c(\omega)]F_b(\omega)|^2 d\omega \tag{23a}$$

$$= \int_{-\infty}^{\omega_a} |[Z_r(\omega) - Z_c(\omega)]F_b(\omega)|^2 d\omega + \int_{\omega_a}^{\omega_b} |[Z_r(\omega) - Z_c(\omega)]F_b(\omega)|^2 d\omega + \int_{\omega_b}^{\infty} |[Z_r(\omega) - Z_c(\omega)]F_b(\omega)|^2 d\omega \tag{23b}$$

where $Z_r(\omega)$ is the impedance given by eqn. (2) for some set of resonator parameters meant to approximate $Z_c(\omega)$ within the known frequency interval. Note that Parseval's Theorem implies convergence of this integral for a physical bunch current distribution. Minimizing $\varepsilon^2$ will result in the best possible fit for an approximate resonator in a least-squares sense. Since $Z_c(\omega)$ is only known on the $[\omega_a, \omega_b]$ interval, we have broken the integral into three parts. The middle integral may be minimized as a first step using any of the well-documented least-squares minimization algorithms such as the Gauss-Newton Method. However, because $Z_c(\omega)$ is not known outside the interval, it is not possible to evaluate the other two integrals in general. Instead, we may define a figure of merit to quantify the impact of the unknown integrals to the overall error. We do this by first by defining the following term:

$$\varepsilon_0^2 = \int_{\omega_a}^{\omega_b} |[Z_r(\omega) - Z_c(\omega)]F_b(\omega)|^2 d\omega, \tag{24a}$$

such that

$$\varepsilon^2 = \varepsilon_0^2 + \int_{-\infty}^{\omega_a} |[Z_r(\omega) - Z_c(\omega)]F_b(\omega)|^2 d\omega + \int_{\omega_b}^{\infty} |[Z_r(\omega) - Z_c(\omega)]F_b(\omega)|^2 d\omega, \tag{24b}$$

where the value of $\varepsilon_0^2$ is a necessary product of the least squares fitting process. Again, we cannot evaluate the unknown integrals exactly, but we may make observations as to what the integrand would have to look like for the error to be non-negligible. Namely, the error is directly proportional to the values of $|Z_r(\omega) - Z_c(\omega)|$ outside $[\omega_a, \omega_b]$. For each frequency point, let

$$Z_c(\omega) = aZ_r(\omega)e^{j\varphi}, \tag{25}$$

where $a$ is some real constant greater than 0 and $0 \leq \varphi \leq 2\pi$. By substituting eqn. (25) into $|Z_r(\omega) - Z_c(\omega)|$ and taking the derivative with respect to $\varphi$ while keeping $a$ static, we observe that the difference magnitude exhibits critical points when $\varphi$ is an integer multiple of $\pi$. By inspection, then, we see that the difference magnitude is at its largest when the phase difference between $Z_c(\omega)$ and $Z_r(\omega)$ is $\pi$, making this the worst-case scenario for a given value of $a$. Substituting eqn. (25) into the two integral terms of eqn. (24b) and setting $\varphi$ equal to $\pi$, we now define the following condition:

$$\varepsilon_0^2 \gg \int_{-\infty}^{\omega_a} (1+a)^2 |Z_r(\omega)F_b(\omega)|^2 d\omega + \int_{\omega_b}^{\infty} (1+a)^2 |Z_r(\omega)F_b(\omega)|^2 d\omega. \tag{26}$$

Here, we make the definition that the above condition is met when

$$p\varepsilon_0^2 \geq \int_{-\infty}^{\omega_a} (1+a)^2 |Z_r(\omega)F_b(\omega)|^2 d\omega + \int_{\omega_b}^{\infty} (1+a)^2 |Z_r(\omega)F_b(\omega)|^2 d\omega, \tag{27}$$

where $p$ is chosen by the one doing the analysis based on their accuracy preference. Nominally, the $\gg$ in eqn. (26) should imply that the RHS of the equation is at least 1 order of magnitude smaller than the LHS, indicating that $p$ should be no more than 0.1. Replacing the inequality with an equal sign and solving for $a$, we obtain

$$a = \sqrt{\frac{p\varepsilon_0^2}{\int_{-\infty}^{\omega_a}|Z_r(\omega)F_b(\omega)|^2 d\omega + \int_{\omega_b}^{\infty}|Z_r(\omega)F_b(\omega)|^2 d\omega} - 1}. \tag{28}$$

$a$, then, becomes our figure of merit for quantifying the error impact of the diverging impedance curves outside the $[\omega_a, \omega_b]$ interval. A larger value of $a$ implies that the difference between the modeled and actual impedances can be larger while still satisfying the error condition of eqn. (27). Conversely, if $a$ is small, the deviation between the true and modelled impedances outside the known frequency interval will have more impact, potentially threatening accuracy. Additionally, eqn. (28) reinforces the idea that the $[\omega_a, \omega_b]$ frequency interval should be chosen such that $F_b(\omega)$ is very small outside this range, as this causes $a$ to increase. A larger $p$ also increases $a$, meaning that it is easier to obtain a certain minimum value of $a$ when the error tolerance is higher. Note that a "good" value of $a$ has not been defined, as this is highly dependent on the use case. Future investigation into the dynamics between $a$ and accuracy will be the topic of future studies.

Finally, we will discuss the evaluation of the integrals in eqn. (28). If the $Z_r(\omega)F_b(\omega)$ product falls off sufficiently quickly, it may be justifiable to evaluate truncated versions of these integrals numerically. Alternatively, one may evaluate the sum of these integrals by invoking the following property:

$$\int_{-\infty}^{\omega_a} |Z_r(\omega)F_b(\omega)|^2 d\omega + \int_{\omega_b}^{\infty} |Z_r(\omega)F_b(\omega)|^2 d\omega = \int_{-\infty}^{\infty} |Z_r(\omega)F_b(\omega)|^2 d\omega - \int_{\omega_a}^{\omega_b} |Z_r(\omega)F_b(\omega)|^2 d\omega \tag{29}$$

The first term on the RHS of eqn. (29) may be evaluated analytically (though somewhat tediously) using the techniques discussed in Section III since $|Z_r(\omega)F_b(\omega)|^2$ and $Z_r(\omega)F_b(\omega)$ have the same poles, only at a higher order. The second term may not be available analytically, but a numerical evaluation in this case would not require truncation, improving accuracy.

## 6. Examples

In this section, we will apply the analysis above to some relevant bunch profiles seen in accelerators. It is typical to fit bunch profiles to the Gaussian function, but it is often more accurate to use distributions other than this. The first distribution chosen for investigation here is the q-Gaussian, which is a generalization of the Gaussian distribution [13]. There are conditions where a beam bunch distribution, particularly the tail portion, is more accurately represented with a q-Gaussian function. This is the case at the LHC for example [14]. The q-Gaussian may be described using the following form:

$$G_q(t) = \left[1 - (1-q)\frac{\beta}{\sigma_{tq}^2} t^2\right]^{\frac{1}{1-q}}, \quad q \neq 1. \tag{30}$$

Where $\sigma_{tq}$ is the q-Gaussian standard deviation, and $q$ and $\beta$ are similar parameters that determine the shape of the q-Gaussian curve. Note we have dropped the amplitude coefficient here, as all the following current distributions will be normalized. For this study, we will adopt the same bunch length parameters as those reported in Table 2 of [14] under the "Final Fit Parameters" field. As such, let $\sigma_{tq} = .227\ ns$, $q = .93$, and $\beta = .4444$ ($\beta$ is not stated explicitly in that work, so this value was determined empirically by examining the curves shown in that work).

The second profile that will be applied is the curve described by the solution to the Haissinki equation, which captures the longitudinal shape of electron bunches in Storage Ring Free Electron Lasers with impedance perturbations [15]. In the case of inductive wake forces, this type of curve can be expressed mathematically as follows using eqn. (21) from [16]:

$$H(t) = \frac{e^{-\left(\frac{1 t^2}{2\sigma_h^2}\right)}}{\sqrt{\frac{\pi}{2}}\left(\frac{S}{a}\right)\left(\coth\left(\frac{S}{2a}\right) + \mathrm{erf}\frac{t}{\sqrt{2}\sigma_h}\right)}. \tag{31}$$

Where $\sigma_h$ is a standard deviation parameter, and $S$ and $a$ determine the skewness of the curve. Here, we let $\sigma_h = \sigma_{tq}$, $S = 2$, and $a = .25$ ($S$ and $a$ are chosen to be these values for demonstrative purposes only; they do not represent anything more).

Consider, somewhat arbitrarily, the lowest-order resonance of a single-gap cavity with a gap length $g$ of 10 cm, a resonant frequency of 500 MHz, a Q of 500, an $R_s$ of 50,000 Ω, and a beam moving at a velocity of .8c. The q-Gaussian and Haissinki distributions, stretched by the finite gap width as per the discussion in the previous section, are visualized with respect to this cavity in Fig. 4. We wish to determine the time-domain voltage response to the excitations of these bunches. Consulting eqn. (5b), $r^+$ and $r^-$ are computed to be $.001j + 1$ and $.001j - 1$ respectively, meaning that the nominal Fourier Transforms, expressed as functions of $\omega$, must be evaluated at $\omega = \omega_r(.001j \pm 1)$. Even though the imaginary parts of the evaluation points are small, we will not neglect them for the sake of generality. To numerically compute the FT values at complex frequencies, we will invoke the following FT property:

$$\check{F}(f(t)e^{j\omega_0 t}) = F_b(\omega - \omega_0). \tag{32}$$

Here, we set $\omega_0 = -.001j\omega_r$ and $\omega = \pm\omega_r$ for the evaluation points we are interested in. Using eqns. (30)-(32), we may find the FTs at the given evaluation points by numerically evaluating the truncated Fourier integral at one point and then taking the complex conjugate to find the FT at the other, which is valid given the smoothness and rapid amplitude fall-off of the distributions.

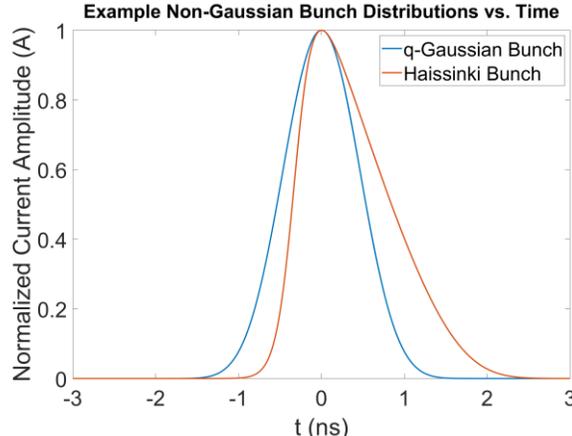

**FIG. 4.** Normalized q-Gaussian and Hassinki bunch current distributions. Note that the Haissinki curve is time-shifted such that its peak occurs at $t = 0$.

Doing this, we find that $F_b(r^\pm) = 4.228 \times 10^{-10} \mp j8.5656 \times 10^{-13}$ for the q-Gaussian distribution and $F_b(r^\pm) = -3.4766 \times 10^{-10} \mp j4.5406 \times 10^{-11}$ for the Haissinki distribution. From eqn. (13), we may use these values to compute the resulting time-domain voltage waveforms within the cavity, which are shown Fig. 5. We may also validate that the curves generated by eqn. (13) are correct by using numerical integration to evaluate the truncated versions of eqn. (1a) directly at each point in time. The results of the direct integration are also shown in Fig. 5. As can be seen in Fig. 5, the eqn. (13) curves show excellent agreement with the direct integration curves, which require many more computations per time point. Examining the curves, we see that differences between the voltage waveforms for the q-Gaussian and Haissinki excitations manifest both in terms of scale and oscillation phase. The Haissinki response amplitude is noticeably smaller and the oscillations of the two responses are slightly phase-shifted. Inspecting the current distributions, this result makes sense; the Haissinki waveform is slightly wider in time than the q-Gaussian, so its Fourier Transform should be narrower, giving rise to a faster voltage decay. Moreover, the asymmetry of the Haissinki bunch would also cause a phase difference to appear since it must be time-shifted for the peak to appear at t = 0. It is true that similar conclusions could be found qualitatively for this example case. However, the purpose of this section is to demonstrate the ease of the technique's application; the method presented here is still effective, at the same level of computational cost, for more general scenarios where such results are far less obvious and precision is key, highlighting the technique's usefulness. Note, for completeness, that the computation outlined here does indeed result in a waveform with a null imaginary part, confirming the physicality of the formulation.

## 7. Conclusion

In this paper, we have used the Cauchy Residue Theorem to develop a general formulation for the evaluation of time-domain resonator voltage waveforms. This formulation incorporates the effect of finite gap lengths and only requires knowledge of the bunch distribution Fourier Transform at one specific evaluation point if this Fourier Transform is holomorphic. The technique has also been extended to multi-modal and non-holomorphic bunch scenarios and accuracy/error considerations have been discussed. This mathematical tool may readily be used to characterize cavity-based beam instabilities in the time-domain without the need to take cumbersome convolution integrals or solve difficult Fourier Transforms analytically. This was shown above, where we easily computed the time-domain waveforms within an example cavity due to two complicated non-Gaussian bunch distributions, demonstrating the power of the method. Moreover, the results using the method were compared to the results of evaluating the corresponding convolution integrals directly and excellent agreement was observed. As such, this technique is highly useful for a wide swath of beam analysis tasks within the accelerator community.

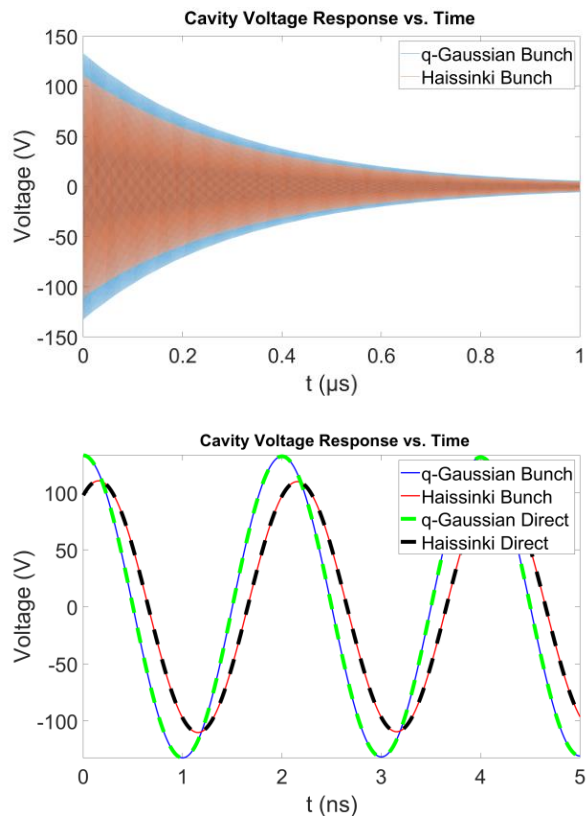

**FIG. 5.** Zoomed-out (top) and zoomed-in (bottom) time-domain cavity voltage waveform due to non-Gaussian bunch current excitations. Dashed lines in the bottom image show the curves generated by numerically evaluating the convolution integrals directly.

11## References

[1] K. Y. Ng, *Physics of Intensity Dependent Beam Instabilities* (World Scientific Publishing, Hackensack, NJ, 2006).

[2] V.G.Vaccaro, CERN ISR-RF/66-35 (1966).

[3] L.Palumbo, V.G.Vaccaro, M.Zobov, Wake fields and impedance, LNF-94-041-P, Sep.1994.

[4] S. Antipov, "Coupled bunch stability with variable filling patterns in PETRA IV", in Proc. IPAC'23, Venezia.

[5] M. Migliorati and L. Palumbo, "Multibunch and Multiparticle Simulation Code with an Alternative Approach to Wakefield Effects," in *Physical Review Special Topics-Accelerators and Beams*, **18**, 031001 (2015).

[6] E. Benedetto et. al., "Simulation study of electron cloud induced instabilities and emittance growth for the CERN Large Hadron Collider proton beam," in in *Physical Review Special Topics-Accelerators and Beams*, **8**, 124402 (2005).

[7] E. Métral, "Fast High-Intensity Single-Bunch Transverse Coherent Instability in Synchrotrons due to a Broad-Band Resonator Impedance", CERN/PS 2001-035 (AE), July 2001.

[8] V. Smaluk, "Beam-Based Impedance Measurement Techniques," in *Proceedings of 58th ICFA Advanced Beam Dynamics Workshop on High Luminosity Circular e+e– Colliders, Daresbury, UK, October, 2016*.

[9] A. Hofmann, "Beam Instabilities", CERN - European Organization for Nuclear Research, Geneva (Switzerland) (1993), in *Proceedings of CAS - CERN Accelerator School: 5th Advanced Accelerator Physics Course*.

[10] A. Chao et. al., *Handbook of Accelerator Physics and Engineering* (World Scientific Publishing, Hackensack, NJ, 2023).

[11] J.W. Brown and R. V. Churchill, *Complex Variables and Applications, Eighth Edition* (McGraw-Hill, New York, NY, 2009).

[12] J. E. Marsden and M. J. Hoffman, *Basic Complex Analysis* (W. H. Freeman and Company, New York, NY, 1999).

[13] P. S. S. Rodrigues and G. A. Giraldi, "Fourier Analysis and q-Gaussian Functions: Analytical and Numerical Results". arXiv:1605.00452v1, (2016).

[14] S. Papadopoulou et. al., "Modelling and Measurements of Bunch Profile at the LHC," in *Proceedings of the 8th International Particle Accelerator Conference, Copenhagen, Denmark, June, 2017*.

[15] C.A. Thomas et. al., "Storage Ring Free Electron Laser Dynamics in Presence of an Auxiliary Harmonic Radio Frequency Cavity," in *Eur. Phys. J. D*, **32**, 83-93 (2005).

[16] Y. Shobuda and K. Hirata, "The Existence of a Static Solution for the Haissinski Equation with Purely Inductive Wake Force," in *Particle Accelerators*, vol. 62, pp. 165-177, (1999).